\documentclass[aps,pra,twocolumn,longbibliography,groupedaddress,showpacs,floatfix]{revtex4-1}
\usepackage{mathtools,amssymb,graphicx,units,xfrac,bm,bbold}
\usepackage{times,mathptmx} 
\usepackage[plainpages=false,
	pdfpagelabels,
	colorlinks=true,
	linkcolor=red,
	urlcolor=blue,
	citecolor=blue,
	pdftitle={Title},
	pdfauthor={},
	pdfdisplaydoctitle=true,
	pdfduplex=DuplexFlipLongEdge]{hyperref}
\usepackage{soul} 
\usepackage[dvipsnames,usenames,svgnames]{xcolor}
\usepackage[normalem]{ulem}
\usepackage{color}
\graphicspath{{./Figures/}} 
\usepackage{etoolbox}
\usepackage{cancel}
\usepackage{soul} 
\usepackage{lipsum}
\usepackage{float}
\usepackage{mathptmx} 
\usepackage{amsmath,empheq}
\usepackage{mathrsfs}
\usepackage{braket}
\usepackage{mhchem}
\usepackage{esvect} 

\usepackage{wasysym,latexsym, amsfonts, amssymb} 

\usepackage{hyperref}
\hypersetup{colorlinks=true, linkcolor=blue, citecolor=blue, urlcolor=blue}

\emergencystretch=\maxdimen
\def\rm#1{\mathrm{#1}}
\def\bf#1{\mathbf{#1}}

\def\rm#1{\mathrm{#1}}

\def\bf#1{\mathbf{#1}}

\begin{document}
\normalem
\title{Superconducting diode effect in a meso-wedge geometry with Abrikosov vortices}
\author{C. A. Aguirre$^{1,2}$
J. Barba-Ortega$^{3}$ 
A. S. de Arruda$^{1}$
and
J. Faúndez$^{4}$ 
}
\affiliation{$^1$ Instituto de F\'isica, Universidade Federal de Mato-Grosso, Cuiab\'a, Brasil.}
\affiliation{$^2$ Escuela Superior de Empresa Ingenier\'ia y Tecnolog\'ia, Bogot\'a, Colombia.}
\affiliation{$^3$ Departamento de F\'isica, Universidad Nacional de Colombia, Bogot\'a, Colombia.}
\affiliation{$^4$ Departamento de Física y Astronomía, Universidad Andrés Bello, Santiago 837-0136, Chile.}
\date{\today}
\begin{abstract}
In this study, we explore the behavior of a superconducting meso-wedge geometry in 3+1 dimensions (three spatial dimensions plus time) subjected to external transport currents at its boundaries and surfaces, as well as external fields applied along the $\hat{z}$-direction. The transport currents are included as two opposite polarities, $\textbf{J}>0$ and $\textbf{J}<0$. Using the generalized time-dependent Ginzburg-Landau theory and considering the order parameter $\kappa$, we focus on two scenarios: a fixed external magnetic field with variable $\kappa$, and fixed $\kappa$ with variable external magnetic field. As a result, under both scenarios, we analyze the voltage-current characteristics of the superconducting meso-wedge, finding that the critical currents differ between polarities, demonstrating the system's non-reciprocity. We further examine the efficiency of the diode as a function of $\kappa$ and the external magnetic field applied. Furthermore, our observations reveal that the current polarity strongly influences the vortex configuration, the parameter $\kappa$, and the applied magnetic field. In particular, the formation of Abrikosov-type vortices exhibits pronounced inhomogeneity depending on the direction of the transport currents. This underscores that the diode effect in the superconducting meso-wedge is intimately associated with the anisotropic nucleation of Abrikosov vortices. Notably, the emergence of polarity-dependent vortex patterns can serve as a distinctive hallmark of the diode effect in these superconducting systems.
\end{abstract}
\date{\today}
\maketitle
\section{Introduction}\label{sec1}
The study of Abrikosov vortices in superconducting systems and their relationship to the diode effect has become a central topic in contemporary condensed matter physics \cite{1,2,3}. Abrikosov vortices arise in type-II superconductors, which are characterized by the Ginzburg–Landau parameter (GLP) $(\kappa>1/\sqrt{2})$. These vortices represent quantized magnetic flux lines that penetrate the superconductor under an applied magnetic field, allowing the coexistence of superconducting and normal-state regions \cite{8,9,10}. Physically, an Abrikosov vortex consists of a core where the superconducting order parameter $\psi$ vanishes, surrounded by circulating super-currents (Meissner currents) that screen the magnetic field. The arrangement and dynamics of these vortices are governed by a balance of long-range attractive and short-range repulsive interactions, strongly influenced by the geometry of the sample and external conditions \cite{a1,a2,a3,a4,a5,a6}.\\\\
A powerful theoretical framework to describe these phenomena is the Ginzburg–Landau theory (GL), which provides a macroscopic description of superconductivity in terms of a complex order parameter $\psi$ and accounts for the interplay between magnetic fields, currents, and the superconducting condensate. In recent years, extensions of the GL theory have enabled the exploration of more complex systems, including mesoscopic superconductors, multi-band and multi-component superconductors, fractional vortices, and topological phases \cite{In5,In6,In7,In8}. Importantly, the interaction of vortices with system geometry can generate rich spatial patterns that critically impact the electromagnetic and transport properties of superconducting materials \cite{In9}. Geometric confinement can alter the mobility, stability, and configurations of the vortex, thus tuning the macroscopic response of the material \cite{Milo1,Milo2,Milo3}.\\\\
A related and rapidly emerging phenomenon is the superconducting diode effect, which refers to the asymmetric response of a superconductor to transport currents of opposite polarities \cite{Diode1,Diode2,Diode3}. This effect manifests itself as a directional dependence in the critical current or resistance, enabling superconducting transport in one direction while suppressing it in the opposite direction. The diode effect is generally associated with symmetry-breaking mechanisms, such as geometric asymmetry, intrinsic material anisotropy, or, in some cases, spin-orbit interactions that break time-reversal symmetry \cite{Diode3}. While the diode effect has been widely studied in Josephson junctions and engineered hetero-structures, its manifestation in mesoscopic superconductors, particularly in systems hosting Abrikosov vortices, remains relatively unexplored. Anisotropy and geometric asymmetry can strongly modify vortex behavior, introducing directionality into the superconducting response and enabling novel functionalities for electronic applications \cite{Diode1,Diode3}. Theoretical approaches such as the London model and numerical simulations based on the time-dependent Ginzburg–Landau equations have been employed to investigate vortex–geometry interactions \cite{Milo1,Milo2}. These studies have revealed mechanisms by which the interplay between vortices and asymmetric boundaries can induce non-reciprocal current transport, providing a pathway to superconducting rectification and new device concepts. Experimental observations have demonstrated the diode effect in a range of systems, including van der Waals hetero-structures without magnetic fields \cite{dio1,dio2,dio3}, thin films of conventional superconductors under weak fields \cite{dio4,dio5,Sharma2024,15}, and twisted tri-layer graphene, where the coexistence of superconductivity and magnetism enables diode-like behavior \cite{16,17}. Non-centrosymmetric superconductors, characterized by broken inversion symmetry and strong spin-orbit coupling, have also emerged as promising platforms \cite{Bav}.\\\\
In this work, we investigate the superconducting diode effect in a meso-wedge-shaped superconductor with broken reflection symmetry along the $\hat{y}$-direction (see Fig.~\ref{Layout}(a) for details). Our main goal is understanding how geometric anisotropy influences these system's critical currents, diode efficiency, and vortex nucleation. Specifically, we analyze the behavior of the superconducting diode effect as a function of the GLP, labeled $\kappa$, and the applied external magnetic field. We provide a detailed explanation of how periodic energy barriers at the system boundaries contribute to the rectification effect, distinguishing this mechanism from that of conventional Josephson-based superconducting diodes \cite{NonVortex}. This paper is organized as follows. Section~\ref{sec2} describes the theoretical framework, including the time-dependent Ginzburg–Landau equations used in our analysis. Section~\ref{sec3} shows the main results from our numerical simulations, including voltage-current characteristics, critical current, diode efficiency, and vortex nucleation varying $\kappa$ or magnetic fields. Finally, we summarize our conclusions and outline future directions in Section~\ref{sec4}.
\section{Theory and Model}\label{sec2}
\begin{figure}[H]
\centering   
\includegraphics[scale=0.42]{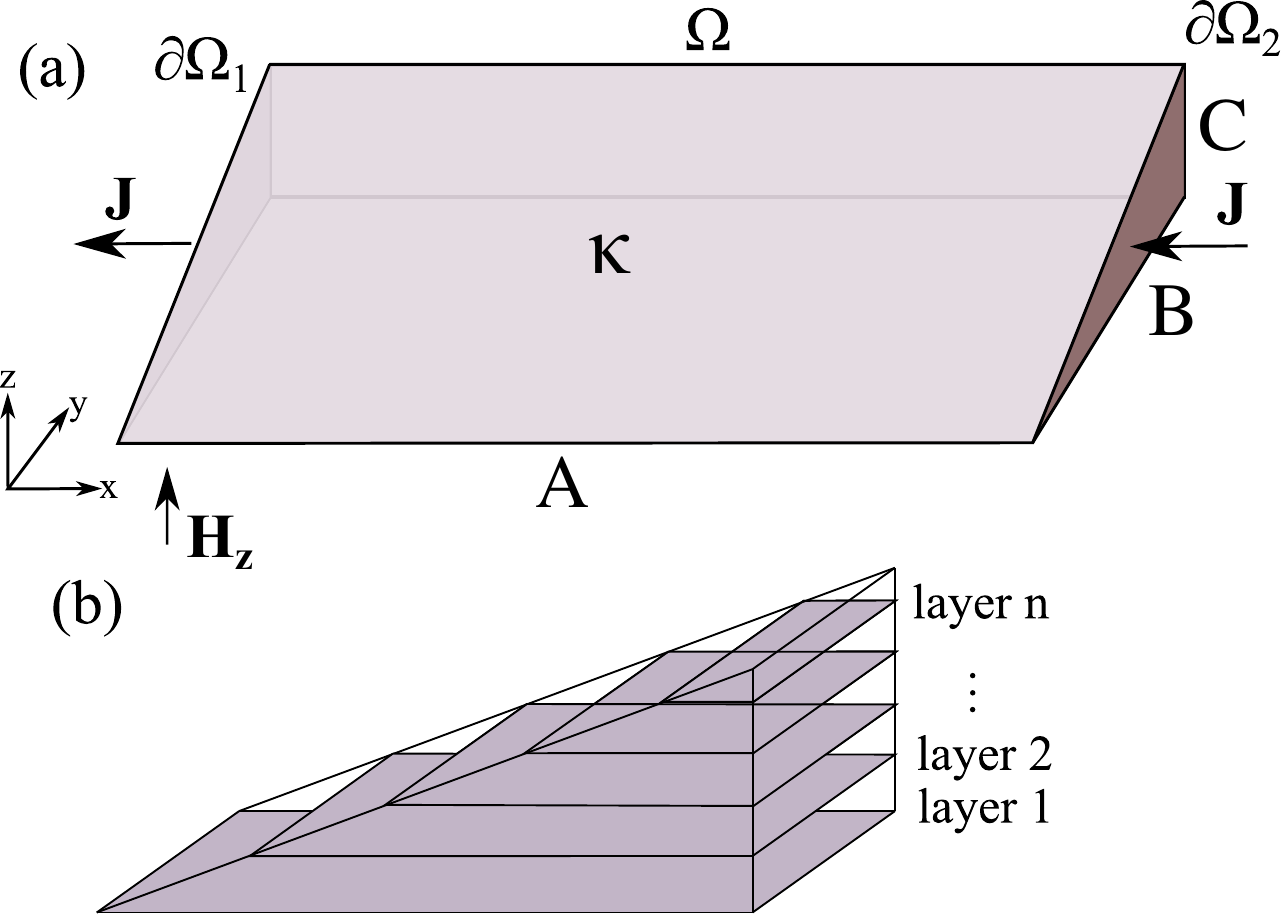}
\caption{(a) Schematic representation of the superconducting meso-wedge geometry. The dimensions of the external numerical mesh are A$=30\xi$, B=C$=15\xi$. (b) Real three-dimensional superconducting sample with meso-wedge geometry and its projection of $n$-layers for the superconducting meso-wedge sample. The inclusion of external/transport currents (${\bf J}$) -in $-\hat{x}$-direction- is given by the lateral faces $\partial \Omega_{i}$, $i=1,2$, the external magnetic field (${ \mathbf{H}=\bf H_{z}}$) is in the $\hat{z}$-direction.}
\label{Layout}
\end{figure}

We studied a real three-dimensional superconducting meso-wedge under a fixed external magnetic field ($\mathbf{H}={\bf H}_{z}$) applied in $\hat{z}$ direction. The geometry of our superconducting meso-wedge is illustrated in Fig. \ref{Layout} (a)-(b). The superconducting meso-wedge fills the domain $\Omega$. The interface between the lateral region and the vacuum is denoted by $\partial\Omega_{i}$, $i=1,2$. The dimension of the numerical sample is A$\times$B$\times$C. With this in mind, an external transport current ($\mathbf{J}$) -in $-\hat{x}$-direction- is applied to the superconducting meso-wedge on the lateral of the geometry, see Fig. \ref{Layout}(a). In this work, we employ the Generalized Time-Dependent Ginzburg-Landau Theory (GTDGL), formulated under the dirty limit and expressed in dimensionless units, as described in Refs.~\cite{In16,In17,In18,In19,In20}.\\
\begin{eqnarray}
\frac{1}{\sqrt{1+\Gamma^{2}|\psi|^{2}}}\bigg[\frac{\partial \psi}{\partial t}+\frac{\Gamma^{2}}{2}\frac{\partial |\psi|^{2}}{\partial t} +\Phi \psi \bigg] \nonumber \\ \nonumber\\
=(i\mbox{\boldmath $\nabla$}+{\bf A})^{2}\psi+\psi(1-|\psi|^2),
\label{EQ1} 
\end{eqnarray}
for the potential vector
\begin{eqnarray}
\frac{\partial{\bf A}}{\partial t}={\bf J}_s-\kappa^2(\mbox{\boldmath $\nabla$}\times\mbox{\boldmath $\nabla$}\times
{\bf A}),
\label{EQ2} 
\end{eqnarray}
where
\begin{eqnarray}
{\bf J}_s = {\rm Re}\left [
\bar{\psi}{(-i\mbox{\boldmath $\nabla$}-{\bf A})}
\psi\right]-\mbox{\boldmath $\nabla$} \Phi.
\label{EQ3}
\end{eqnarray}
In conjunction with the continuity equation, which also adopted the Coulomb gauge $\nabla \cdot \mathbf{A}= 0$ and Maxwell's first law, the expression for the scalar potential ($\Phi$) is obtained as a Poisson time-dependent equation, which is given by:
\begin{eqnarray}
\mbox{\boldmath $\nabla$}^{2} \Phi= \frac{\partial \rho}{\partial t}=-\nabla \cdot \mathbf{J}.
\label{EQ4}
\end{eqnarray}
With this, the voltage is calculated for a given applied current, and with Eq. \ref{EQ3}, the voltage at each mesh point can be calculated. The Eqs.~(\ref{EQ1})-(\ref{EQ4}) are solved in a self-consistent approach. The Neumann boundary conditions for the potential/external current are $\hat{\boldsymbol{n}} \cdot \nabla \boldsymbol{\Phi} = -\boldsymbol{J}$ in sections with external current ($\partial \Omega_{i}$, $i=1,2$) and $\hat{\boldsymbol{n}} \cdot \nabla \boldsymbol{\Phi} = 0$ in the other sections, with $\hat{\boldsymbol{n}}$ being a surface normal vector. In addition, $\boldsymbol{J}_{s}$ corresponds to the total superconducting current density (Meissner plus external/transport). In Eqs.~(\ref{EQ1}), (\ref{EQ2}), and (\ref{EQ3}), dimensionless units were introduced as follows: the order parameter $\psi$ is in units of $\psi_{\infty} = \sqrt{-\alpha/\beta}$ (the order parameter at the Meissner-Oschenfeld state), where $\alpha$ and $\beta$ are two phenomenological constants; $\boldsymbol{H}_{1}$ is the first critical field (Meissner-Oschenfeld field); lengths are in units of the coherence length $\xi$; time is in units of the Ginzburg-Landau characteristic time $t_{GL} = \pi\hbar / (8K_B T_c)$; fields are in units of $\boldsymbol{H}_{c2}$, where $\boldsymbol{H}_{c2}$ is the bulk second critical field; the vector potential $\boldsymbol{A}$ is in units of $\xi \boldsymbol{H}_{c2}$; $\kappa = \lambda / \xi$ is the $\mathbf{GLP}$, which describes the type of superconductor as a function of the spatial variation of the order parameter $\psi$ and the penetration of the magnetic field into the sample and $\Gamma = 10$. In addition, we use the triple convergence rule for time~\cite{Nadeem2023,Aguirre2024}.
\begin{eqnarray}
dt_{1} = \frac{a\eta}{4\sqrt{1+\Gamma^{2}}},\quad dt_{2} = \frac{a\beta}{4\kappa^{2}},\quad dt_{3} = \frac{a\nu}{4\zeta^{2}},
\end{eqnarray}
and
\begin{eqnarray}
\Delta t \leq min \{dt_{1},dt_{2},dt_{3}\},\quad a^2 = \frac{2}{\frac{1}{\delta x^2} + \frac{1}{\delta y^2}+\frac{1}{\delta z^2}}.
\end{eqnarray}
For numerical calculations, we use the mesh size $\delta x = \delta y = \delta z = 0.1$, the values of the constants: $\eta=5.79$, $\beta=1.0$, $\zeta=0.50$, $\kappa$ will be variable in a section of this manuscript, and $\nu=0.03$ \cite{In20}. For tolerance in convergence of the order parameter $\psi$, we employ $\epsilon=1.0^{-9}$, and the errors are of order $O(\Delta x)^{2}$ for space and time. For boundary conditions of the order parameter, we employ Robin's boundary condition ${\bf n} \cdot (i\mbox{\boldmath $\nabla$} + {\bf A}) \psi = -i\psi/b$,  with ${\bf n}$ being a surface normal vector and $b$, the de-Gennes extrapolation parameter and we have taken $b\rightarrow \infty$ in the lateral contacts and $b=0$ in the rest of the sample. With this, as shown in Fig. \ref{Layout}(a), the dimensions of the external numerical mesh are $A=30\xi$, $B=C=15\xi$, and in Fig. \ref{Layout}(b), the superconducting meso-wedge has $n$-layers, where C=$n\xi$.
\section{Numerical Results}\label{sec3}
\begin{figure}
    \centering
    \includegraphics[width=1.01
    \linewidth]{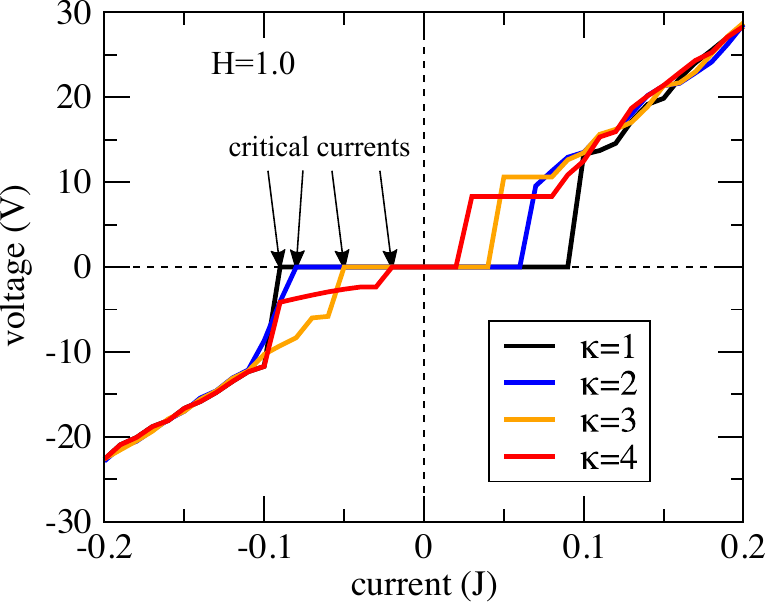}
    \caption{Voltage response ($V$) as a function of the transport current ($\mathbf{J}$) for a fixed external magnetic field applied along the $z$-direction, $\mathbf{H}_{z} = 1.0$, and for both current polarities ($\mathbf{J} > 0$ and $\mathbf{J} < 0$). The results are shown for different values of the Ginzburg–Landau parameter, $\kappa$. The geometry of the superconducting meso-wedge is fixed with dimensions $A = 30\xi$ and $B = C = 15\xi$.}
    \label{figV1}
\end{figure}

\begin{figure}
    \centering
    \includegraphics[width=1.01\linewidth]{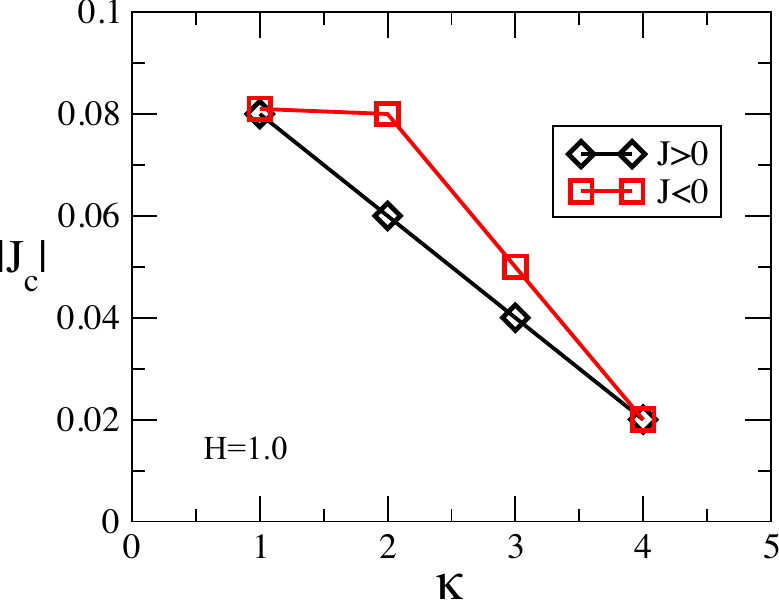}
    \caption{Magnitude of the first critical current ($\mathbf{J}_{c}$) for both polarities of the transport current ($\mathbf{J_{c}}$), plotted as a function of Ginzburg-Landau  parameter, $\kappa$. Results are shown for fixed values of the external magnetic field, $\mathbf{H}=1.0$, in the superconducting meso-wedge sample.}
    \label{Jc1}
\end{figure}

\begin{figure}
    \centering
    \includegraphics[width=1.01\linewidth]{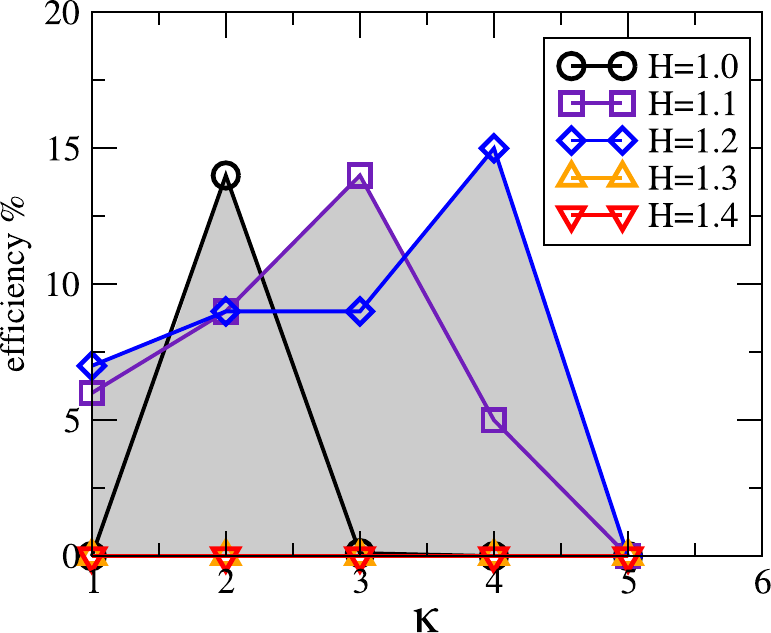}
    \caption{Diode efficiency $\gamma_{d}(\mathbf{H})$ as a function of the Ginzburg–Landau parameter $\kappa$, for fixed magnetic field values $\mathbf{H} \in [1.0,\, 1.4]$ (in steps of $0.1$), in a mesoscopic superconducting wedge acting as a potential diode. As the $\mathbf{H}$ increases, the $\gamma_{d}(\mathbf{H})$ decreases, and for sufficiently large values of $\mathbf{H}$, the diode effect vanishes, i.e., $\gamma_{d}(\mathbf{H}) \rightarrow 0$ for all $\kappa$.}
    \label{Eff1}
\end{figure}

\begin{figure}
    \centering
    \includegraphics[width=1.01\linewidth]{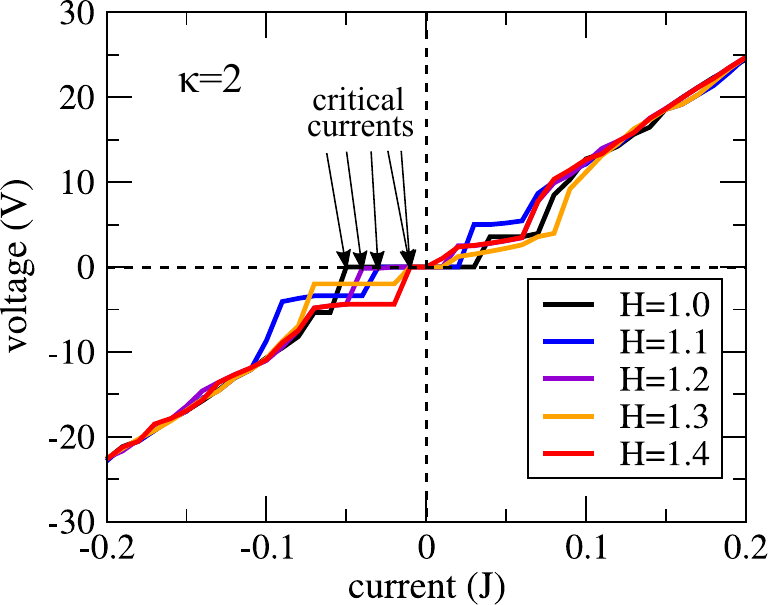}
    \caption{Voltage $V$ as a function of the external current $\mathbf{J}$ for a fixed Ginzburg–Landau parameter $\kappa = 2.0$, considering both current polarities ($\mathbf{J} > 0$ and $\mathbf{J} < 0$), and for different values of the external magnetic field $\mathbf{H} \in [1.0,\, 1.4]$ (in steps of $0.1$). The geometry of the superconducting meso-wedge sample is defined by A$=30\xi$ and B=C$= 15\xi$.}
    \label{figV2}
\end{figure}

\begin{figure}
    \centering
    \includegraphics[width=1.01\linewidth]{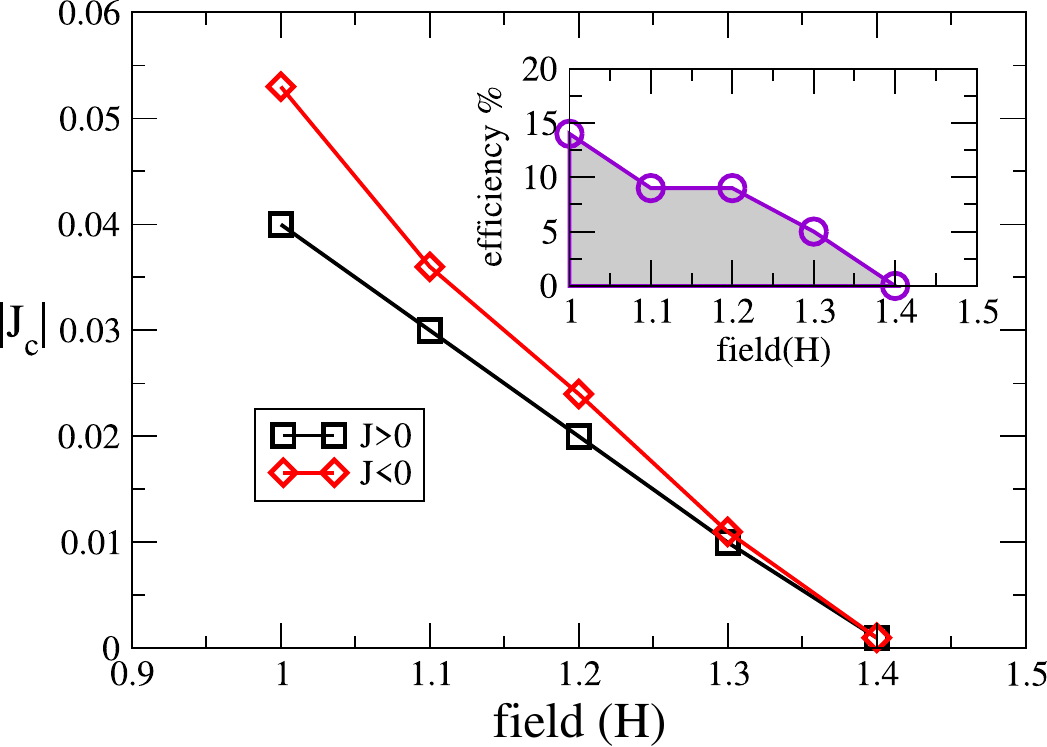}
    \caption{Magnitude of the first critical currents $\mathbf{J}_{c}$, for both polarities, as a function of the fixed external magnetic field $\mathbf{H}$ for a fixed Ginzburg-Landau parameter ($\kappa=2.0$) in the superconducting meso-wedge sample. Inset: efficiency value ($\gamma_{d}(\mathbf{H})\time 100\%$), for both polarities ($\mathbf{J}>0$ and $\mathbf{J}>0$), as a function of the fixed external magnetic field $\mathbf{H}$ for a fixed $\kappa=2.0$ in the superconducting meso-wedge sample.}
    \label{Jc2}
\end{figure}

This section presents the numerical results obtained for the superconducting meso-wedge. To this end, we begin by showing in Fig.~\ref{figV1} the voltage response ($V$) as a function of the externally applied transport current ($\mathbf{J}$) for different values of the GLP ($\kappa$) and the fixed magnetic field, $\mathbf{H_{z}}=1.0$. The values of $\kappa$ considered are all within the type-II superconducting regime and are ordered in increasing magnitude. As $\kappa$ increases, we observe a reduction in the critical current. This behavior arises from an increase in penetration depth, which modifies the slope of the voltage-current curve and leads to the emergence of a transient resistive state. This state is associated with the nucleation and motion of vortices inside the superconducting meso-wedge. Moreover, the transport current along the $-\hat{x}$-direction generates a preferred direction for vortex movement, driven by the Lorentz force \cite{Milo2}. The resulting voltage jumps—reminiscent of the Shapiro steps~\cite{Shapiro}—are related to the maximum velocity of the vortices, given by $v^* = V/B$, where $B$ is the magnetic flux density. The vortex velocity is also influenced by vortex–vortex interactions, which depend on the Meissner current's circulation. These interactions can either enhance or suppress the vortex mobility. Therefore, we associate the mechanical rigidity of the superconducting vortex lattice with the parameter $\kappa$ (here referred to as the GLP): larger values of $\kappa$ reduce the lattice stiffness, facilitating vortex entry at lower values of the external transport current.\\\\
Fig.~\ref{Jc1} shows the magnitude of the critical currents ($\mathbf{J_{c}}$) -the onset of resistive states- as a function of $\kappa$ for both the polarities of $\mathbf{J}$ and fixed $\mathbf{H_{z}}=1.0$. For $\kappa = 1.0$, only a slight difference is observed between $\mathbf{J_{c}}'s$, which may be attributed to the high rigidity of the superconducting condensate, where the strong coupling between Cooper pairs suppresses the vortex dynamics. In contrast, for $\kappa = 2.0$, a more pronounced asymmetry emerges, suggesting that a lower rigidity of the vortex lattice facilitates vortex motion and phonon-like oscillations. This asymmetry increases with $\kappa$ up to a certain value, after which it decreases for higher values of $\kappa$. Beyond the $\mathbf{J_{c}}$ threshold, the mesoscopic superconducting wedge exhibits a nearly Ohmic (linear) response. Importantly, the observed difference between the $\mathbf{J_{c}}'s$ for the positive and negative polarities of $\mathbf{J}$ confirms the presence of the superconducting diode effect. This asymmetry reflects the breakup of the spatial inversion symmetry in the superconducting meso-wedge \cite{Shapiro}.\\\\
We now proceed to characterize the diode effect quantitatively. To this end, it is useful to define a signed efficiency parameter, following the approach introduced in Ref.~\cite{Banerjee2024} (and references therein), which quantifies the degree of rectification by measuring the asymmetry between the critical currents for opposite current polarities. This efficiency parameter is defined as:\\
\begin{equation}
    \gamma_{d}(\mathbf{H})=\frac{|J^{+}_{c}(\mathbf{H})-|J^{-}_{c}(\mathbf{H})||}{J^{+}_{c}(\mathbf{H})+|J^{-}_{c}(\mathbf{H})|}\times 100,
\end{equation}\\
The values of the $\mathbf{J_{c}}'s$ used in the evaluation of the diode efficiency are extracted from the results shown in Fig.~\ref{figV1} and Fig.~\ref{Jc1}, where the onset of resistive states for both polarities of the applied $\mathbf{J}$ is determined. Using this information, in Fig.~\ref{Eff1}, we present the efficiency of the diode as a function of $\kappa$, for several fixed values $\mathbf{H} \in [1.0, 1.4]$ (in steps of $0.1$). For the lowest considered $\mathbf{H} = 1.0$, we observe that the efficiency reaches its maximum at $\kappa = 2.0$. Around this point, the diode effect is most pronounced, but as the $\kappa$ parameter increases, the efficiency decreases, exhibiting a non-monotonic behavior. This indicates an optimal range of $\kappa$ values where the asymmetry between $\mathbf{J_{c}}'s$ is maximized.  As we increase the external value of $\mathbf{H}$, the position of this maximum efficiency shifts to higher values of $\kappa$. This behavior suggests that stronger $\mathbf{H}$'s tend to favor rectification in samples with reduced superconducting rigidity (larger $\kappa$), possibly due to enhanced vortex mobility and a softer vortex lattice. Nevertheless, regardless of the $\mathbf{H}$ strength or the precise value of the $\kappa$ parameter, the efficiency remains bounded and does not exceed approximately 15\%. Furthermore, for very low or very high values of $\kappa$, the efficiency of the diode tends to vanish, as the difference between the $\mathbf{J_{c}}'s$ for positive and negative $\mathbf{J}$ directions becomes negligible. This suppression of the diode effect occurs even though the meso-wedge geometry breaks the spatial inversion symmetry of the system \cite{Bav,In8}. This highlights the crucial role played by the interplay between superconducting rigidity and vortex dynamics in enabling rectification.\\\\
In Fig.~\ref{figV2}, we illustrate $V$ as a function of  $\mathbf{J}$ for a fixed value $\kappa = 2.0$, and for several values of $\mathbf{H} \in [1.0,\, 1.4]$ (in steps of $0.1$). The results are shown for both polarities: $\mathbf{J} > 0$ and $\mathbf{J} < 0$. As $\mathbf{H}$ increases, vortex nucleation and penetration into the superconducting region become more favorable, leading to an earlier onset of the resistive state. Our results confirm this: $\mathbf{J_{c}}$ decreases with increasing $\mathbf{H}$, and the dissipation threshold shifts to lower values of $\mathbf{J}$. Moreover, a clear asymmetry is observed between $\mathbf{J_{c}}$ for opposite polarities of $\mathbf{J}$, indicating the presence of the superconducting diode effect. This asymmetry is more pronounced at lower $\mathbf{H}$'s and progressively weakens as $\mathbf{H}$ increases, consistent with a reduction in diode efficiency. Compared with the results in Fig.~\ref{figV1}, it is evident that the $\mathbf{J_{c}}$'s  are systematically lower due to the enhanced vortex dynamics at higher $\mathbf{H}$ strengths. Beyond $\mathbf{J_{c}}$, following the initial jump $V$ associated with vortex entry, the system enters a regime characterized by an approximately Ohmic (linear) response \cite{In9}. These observations suggest that the diode efficiency as a function of $\kappa$ is likely to exhibit a non-monotonic and nonlinear dependence for varying $\mathbf{H}$ strengths, as illustrated in the case of $\kappa = 2.0$.\\\\
With the previous results, in Fig.~\ref{Jc2}, we present the first $\mathbf{J_{c}}$ values as a function of $\mathbf{H}$, for a fixed value of $\kappa=2.0$. The selected  $\mathbf{H}$'s are all above the first critical field $\mathbf{H}_{1}$, where vortices are expected to nucleate and penetrate the superconductor. In this regime, the vortex configurations can differ between $\mathbf{J}>0$ and $\mathbf{J}<0$ polarities, potentially leading to distinct $\mathbf{J_{c}}$ values—a key signature of the diode effect \cite{Milo1,Milo2,Milo3}. The results show a non-monotonic dependence of the $\mathbf{J_{c}}$ on the $\mathbf{H}$. Notably, the largest asymmetry occurs at $\mathbf{H} = 1.0$, while the $\mathbf{J_{c}}$'s converge and become equal at $\mathbf{H} = 1.4$ for $\kappa = 2.0$. In between, the $\mathbf{J_{c}}$'s exhibit alternating increases and decreases, indicating a complex interplay between vortex dynamics and $\mathbf{H}$ strength. This behavior is consistent with the V-$\mathbf{J}$ characteristics shown in Fig.~\ref{figV2}. It suggests that the diode efficiency, as a function of $\kappa$, is expected to exhibit oscillatory or non-monotonic behavior across different values of the $\mathbf{H}$.\\\\
In the inset of Fig.~\ref{Jc2}, we show $\gamma_{d}(\mathbf{H})$ for a fixed $\kappa = 2.0$, as a function of $\mathbf{H} \in [1.0,\, 1.4]$ (in steps of $0.1$). These results are consistent with the analysis of the $\mathbf{J_{c}}$'s discussed in the previous paragraphs. In particular, $\gamma_{d}(\mathbf{H})$ exhibits an oscillatory behavior with varying $\mathbf{H}$, reaching a maximum at $\mathbf{H} = 1.0$ and decreasing for larger values of $\mathbf{H}$. Interestingly, despite the presence of vortices in the sample, expected for all selected values $\mathbf{H}$ above $\mathbf{H}_{1}$, $\gamma_{d}(\mathbf{H})$ remains non-zero. This suggests that the presence of vortices does not completely suppress the rectification effect and that specific vortex configurations can still lead to a measurable asymmetry between positive and negative $\mathbf{J_{c}}$'s. These findings indicate that $\gamma_{d}(\mathbf{H})$ of the superconducting meso-wedge geometry is highly sensitive to the interplay between the external $\mathbf{H}$, $\kappa$, and the geometric asymmetry of the sample. Importantly, this result supports the idea that the diode effect can arise solely from spatial symmetry breaking, without the need for Josephson junctions, as has been commonly proposed in previous studies~\cite{dio3,dio4}.\\\\
With this, we conclude the presentation and analysis of the electronic transport properties for the superconducting meso-wedge. However, the observed asymmetries in the critical currents and diode efficiency suggest that the underlying vortex configurations play a central role in enabling the rectification effect. To explore this hypothesis further, we examine the spatial distribution of the Cooper pair density in the following section, which provides insight into the vortex dynamics and their correlation with the diode effect.
\subsection{Vortex configuration}
\begin{figure*}
\centering
    \centering \includegraphics[width=1.0\linewidth]{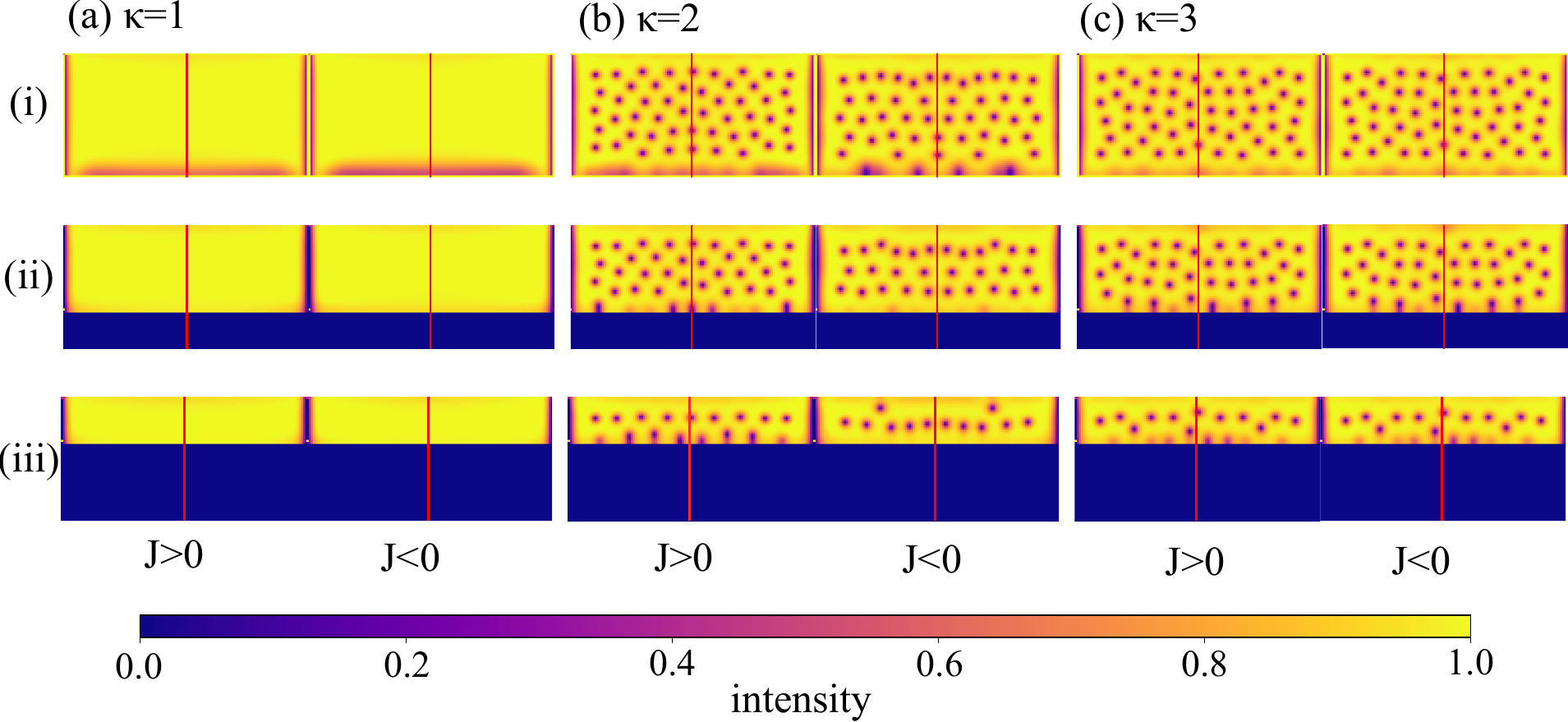}
    \caption{Cooper pair density $|\psi|^2$ for different values of the Ginzburg–Landau parameter $\kappa$. Panel (a): $\kappa = 1.0$; panel (b): $\kappa = 2.0$; and panel (c): $\kappa = 3.0$. The applied transport current is fixed at $\mathbf{J} = 0.15$, and the external magnetic field is set to $\mathbf{H} = 1.0$. The vortex states are projected onto three different layers: (i) $n = 1$, (ii) $n = 4$, and (iii) $n = 10$. The color bar indicates the intensity of $|\psi|^{2}$ across the sample.}
    \label{figVorCn1}
\end{figure*}

\begin{figure*}
\centering
    \centering \includegraphics[width=1.0\linewidth]{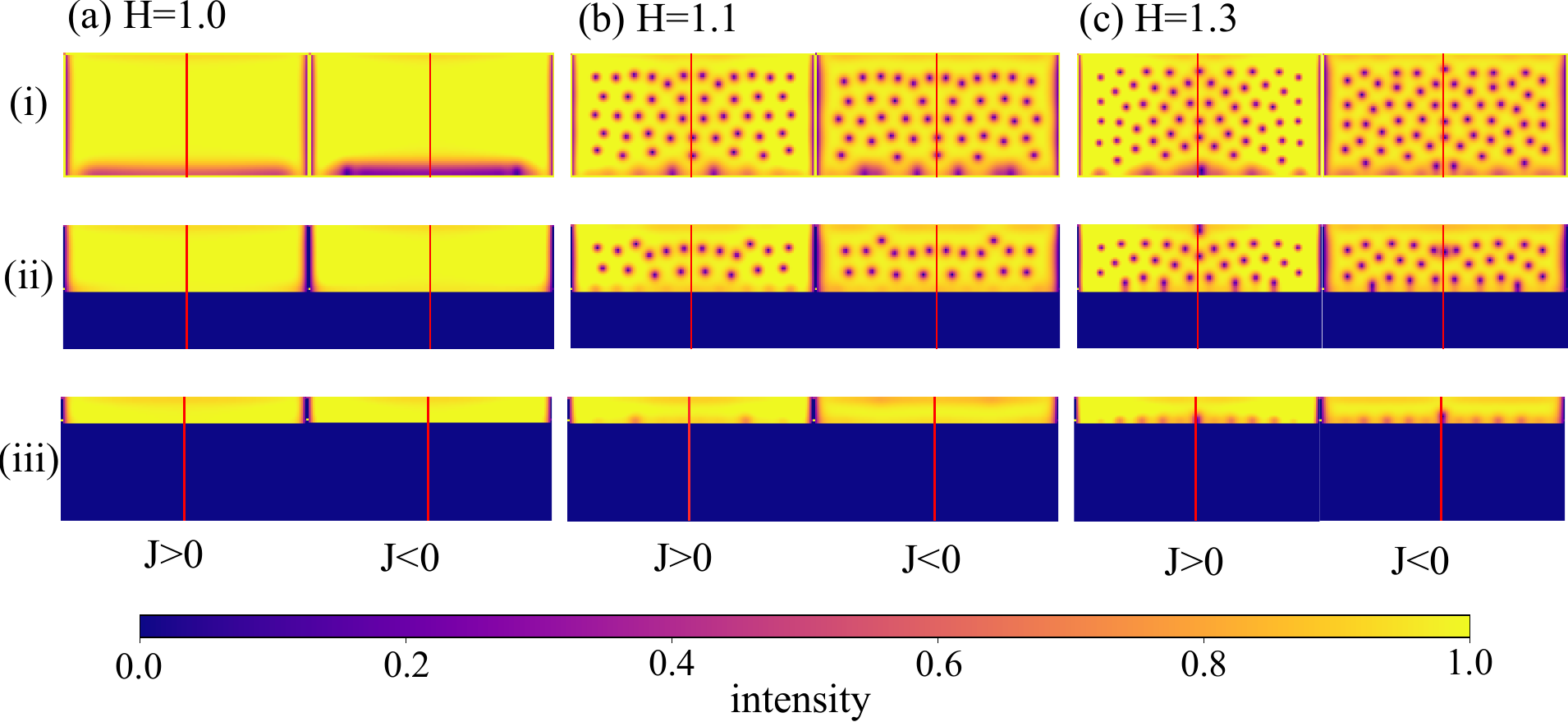}\caption{Cooper pair density $|\psi|^2$ for different values of the external magnetic field $\mathbf{H}$. Panel (a): $\mathbf{H} = 1.0$; panel (b): $\mathbf{H} = 1.1$; and panel (c): $\mathbf{H} = 1.3$. The transport current is fixed at $\mathbf{J} = 0.015$, and the Ginzburg–Landau parameter is set to $\kappa = 2.0$. The density is projected for three selected layers of the superconducting meso-wedge: $n = 1$, $n = 7$, and $n = 13$. The color bar indicates the intensity of $|\psi|^{2}$in each region of the sample.} 
    \label{figVorCn2}
\end{figure*}

The order parameter $\psi$ in Ginzburg–Landau theory represents the macroscopic wave function of the superconducting state, and its squared modulus $|\psi|^2$ corresponds to the local density of Cooper pairs. Spatial variations in $|\psi|^2$ can, therefore, be used to visualize the presence and distribution of vortices, which appear as localized regions of the suppressed Cooper pair density. We now focus on the spatial distribution of the order parameter to investigate the role of vortex configurations in the emergence of the diode effect. Figs~\ref{figVorCn1}(a)–(c) display the Cooper pair density $|\psi|^2$ for a fixed $\mathbf{H}$ and several values of $\kappa$, specifically $\kappa = 1.0$, $2.0$ and $3.0$. The plots are shown for three selected layers of the superconducting meso-wedge, chosen to illustrate differences in material thickness and vortex behavior. These particular layers were selected because the lowest layer exhibits vortex entry. In contrast, vortex penetration is either suppressed or not visible in the higher layers due to the larger amount of material and the presence of a stronger surface energy barrier. In such regions, significantly stronger $\mathbf{H}$ would be required to observe vortex nucleation and entry, as discussed in~\cite{Aguirre2024} and references therein.\\\\
In general, the superconducting state in the upper sections of the superconducting meso-wedge can be partially suppressed due to geometry-induced variations in $\mathbf{J}$ and $\mathbf{H}$ screening. This leads to the formation of periodic energy barriers that modulate vortex entry and dynamics. Our simulations show that vortices tend to nucleate at the boundary where the sample is thinner. This is energetically favorable due to the reduced surface energy and weaker screening $\mathbf{J}$ in that region. This boundary-induced asymmetry allows vortices to enter more easily from one side of the sample, especially under finite $\mathbf{J}$. Both geometric and energetic factors govern the interaction between the vortices and the material boundaries. The magnetic penetration depth increases for higher values of $\kappa$, resulting in more extended Meissner $\mathbf{J}$ and a softer vortex–boundary interaction. Consequently, the effective energy barrier for vortex entry is lowered, making the superconducting state in these regions more resilient to disruption, even under $\mathbf{H}$ and $\mathbf{J}$. However, in thicker sections of the wedge (higher layers), the energy barrier is stronger, primarily due to the larger effective thickness and enhanced surface screening. This creates a so-called geometric barrier, well documented in mesoscopic superconductors~\cite{Zeldov1994, Benkraouda1996}, which delays vortex entry until the local field exceeds a threshold.\\\\
The presence of vortices near the boundaries modifies the local superconducting properties by introducing periodic regions of suppressed $|\psi|^{2}$. These vortex cores act as anchor centers and form an effective periodic potential landscape, influencing the flow of $\mathbf{J}$. This modulation leads to localized $\mathbf{J}$ density amplification near the edges, reinforcing the asymmetry introduced by the superconducting meso-wedge. We compare the vortex configuration for opposite $\mathbf{J}$ reversal to examine how this asymmetry manifests under $\mathbf{J}$ polarities. Due to the explicit breaking of spatial symmetry concerning $\hat{x}$-direction, we observe that the vortex arrangements are not identical. Even though the magnitude of the transport $\mathbf{J}$ is the same, the effective Lorentz force acting on the vortices—and how they interact with boundaries and pinning centers—differs depending on the transport $\mathbf{J}$ direction ($\hat{x}$ and -$\hat{x}$). This difference constitutes the first microscopic evidence of a diode-like response in the superconducting state with vortices present. It is in agreement with the observed asymmetry in the critical $\mathbf{J}$  values reported in Figs.~\ref{figV1} and \ref{Jc1}. Furthermore, the vortex dynamics are shaped by a rich interplay of forces: vortex–vortex repulsion (which scales logarithmically in two dimensions), interaction with energy barriers (determined by material geometry and superconducting parameters), and coupling to the Meissner and transport $\mathbf{J}$'s. Because of this complexity, the resulting vortex configurations are highly non-monotonic and do not resemble the ideal Abrikosov triangular lattice~\cite{Milo1,Milo2}. Notably, these configurations are sensitive not only to $\kappa$ and $\mathbf{H}$ but also to the polarity of the transport $\mathbf{J}$, reinforcing the notion that rectification effects can emerge purely from geometric and dynamical asymmetries, even in the absence of Josephson junctions.
In Figs.~\ref{figVorCn2}(a)–(c), we show  $|\psi|^{2}$ for the superconducting meso-wedge under different values of $\mathbf{H} = 1.0,\ 1.1,\ 1.3$, while keeping $\kappa = 2.0$. This value of $\kappa$ corresponds to the case where the diode efficiency reaches its maximum, specifically at $\mathbf{H} = 1.0$, as discussed previously. As in the previous analysis (see Fig.~\ref{figVorCn1}), the vortex configuration deviates from the conventional Abrikosov triangular lattice. This deviation is attributed to the combined influence of vortex–vortex interactions, boundary-induced energy barriers, and the Lorentz force exerted by the transport $\mathbf{J}$. \\\\
In this case, we observe that the variation in vortex configurations between the two $\mathbf{J}$ polarities is more pronounced, particularly for $\mathbf{H} = 1.0$, where the diode efficiency is highest. Interestingly, as $\mathbf{H}$ increases, the differences in vortex configurations between positive and negative $\mathbf{J}$ polarities diminish, leading to more symmetric and homogeneous states. This observation provides a microscopic basis for reducing diode efficiency at higher fields: the rectification effect is effectively canceled when the vortex distributions become identical for both polarities. However, near the sample boundaries, residual asymmetries remain visible even at higher fields, especially due to the geometric inhomogeneity across layers, which leads to a different number of vortices (i.e., different vorticity) depending on the layer and the direction of the transport $\mathbf{J}$. Finally, as $\mathbf{H}$ increases, we observe the emergence of a stronger energy barrier at the sample boundary. This sharper contrast between superconducting and normal regions in $|\psi|^{2}$ forms a periodic energy wall near the edges. From an energetic perspective, this boundary structure modulates the vortex dynamics and is directly responsible for the inhomogeneous critical $\mathbf{J}$ observed in the previous Figs. \ref{figV2} and \ref{Jc2}.\\\\
Having established the theoretical framework, we now focus on the experimental observations that validate and illustrate the physical behavior of the superconducting meso-wedge.
\subsection{Discussion and experimental evidences}
There is increasing experimental support for the presence and role of vortices in the emergence of the superconducting diode effect. Although imaging and resolving vortex configurations under specific values of the external magnetic field $\mathbf{H}$ and applied current $\mathbf{J}$ can be technically challenging, recent advances—particularly the development of nanoscale SQUID-on-tip (SOT) microscopy have enabled direct measurements of local magnetic flux and vortex distributions in superconducting systems. One of the most detailed experimental studies in this context is presented by A. Gutfreund \textit{et al.}~\cite{Exp1}, who investigate Nb/EuS (S/F) bilayers and measure the voltage $V$ as a function of the external applied current $\mathbf{J}$ for a fixed $\kappa$. Remarkably, their data exhibit distinct Shapiro steps—quantized voltage plateaus that result from phase locking between the time-dependent superconducting order parameter and an external frequency scale, such as the motion of vortices or internal Josephson oscillations. The appearance of these steps in their measurements closely resembles our theoretical predictions shown in Figs.~\ref{figV1} and \ref{figV2}, where similar features arise from dynamic vortex entry and collective motion under increasing $\mathbf{J}$. In addition, Gutfreund \textit{et al.} report pronounced asymmetries in the critical $\mathbf{J}$ values for opposite polarities of $\mathbf{J}$ ($\mathbf{J}>0$ and $\mathbf{J}<0$), consistent with the diode-like behavior predicted in our model (Figs.~\ref{Jc1} and \ref{Jc2}). Importantly, their imaging of vortex configurations reveals non-trivial spatial arrangements: instead of forming regular Abrikosov triangular lattices, vortices are found to align along the sample boundaries or exhibit irregular clustering patterns. These features are shaped by the combined effects of geometric confinement, surface energy barriers, and the Lorentz force acting on vortices under transport $\mathbf{J}$'s. Strikingly similar patterns emerge in our theoretical $|\psi|^2$ maps (Figs.~\ref{figVorCn1} and \ref{figVorCn2}), particularly near asymmetric boundaries, where vortex nucleation and trapping are strongly geometry-dependent. Such \textit{vortex asymmetries} is critical for manifesting the diode effect: rectification arises when the vortex dynamics—specifically their nucleation sites, mobility, and paths—differ under current reversal. No net diode effect would be observed if vortex motion were symmetric for both directions of $\mathbf{J}$. Therefore, the close agreement between vortex configurations and critical $\mathbf{J}$ asymmetries in both our theoretical model and the experimental findings of Gutfreund \textit{et al.} provides strong evidence that vortex-mediated mechanisms are central to the non-reciprocal transport observed in superconducting diode systems.\\\\
Castellani \textit{et al.}~\cite{Exp2} explores the diode effect in superconducting niobium nitride micro-bridges in a complementary work. They measure the critical $\mathbf{J}$'s and diode efficiency. Their results reveal a peak in efficiency at specific $\mathbf{H}$ and $\mathbf{J}$ combinations, which follows a functional dependence remarkably similar to that obtained in our simulations (Fig.~\ref{Eff1} and Fig.~\ref{Jc2}, inset). This supports the hypothesis that diode efficiency is maximized when vortex configurations are strongly asymmetric and minimized when the system approaches dynamical symmetry. Also, Taras \textit{et al.}~\cite{Exp3} investigate the superconducting diode effect from the perspective of \textit{nonreciprocity induced by spatial symmetry breaking} in a conventional Nb superconductor. Notably, their experiments demonstrate that even without an external magnetic field, geometric asymmetries in the sample can lead to rectification of the applied $\mathbf{J}$. They show that the difference between forward and reverse critical $\mathbf{J}$'s persists up to zero temperature, an observation that matches the predictions of our theoretical framework in the limit of low thermal fluctuations and dominant geometric effects.\\\\
These experimental studies validate the formalism and results presented in this work. The convergence between theory and experiment—regarding vortex distributions, critical current asymmetries, Shapiro steps, and diode efficiency—highlights the key role of mesoscopic geometry and vortex dynamics in the superconducting diode effect. Our results suggest that nonreciprocal superconducting behavior does not necessarily require Josephson junctions or spin-orbit coupling, but can emerge naturally from the interplay between vortex physics and broken spatial symmetry.
\section{Conclusions}\label{sec4}
In this work, we have theoretically investigated the superconducting diode effect in a superconducting meso-wedge, focusing on the influence of the Ginzburg–Landau parameter $\kappa$ and the external magnetic field $\mathbf{H}$ on the resistive state and vortex dynamics. By computing the critical current $\mathbf{J}_c$ for both transport current $\mathbf{J}$ polarities ($\mathbf{J} > 0$ and $\mathbf{J} < 0$), we identified the emergence of a diode effect characterized by asymmetric transport, quantified via an efficiency parameter $\gamma_{d}(\mathbf{H})$. Notably, $\gamma_{d}(\mathbf{H})$ exhibits a non-monotonic dependence on $\kappa$, reaching a maximum at intermediate values and vanishing for larger $\kappa$, suggesting a relationship with the rigidity of the superconducting condensate and the interplay between vortex mobility and energy barriers.\\\\
Our simulations reveal that the diode effect persists even in Abrikosov vortices and does not rely on Josephson junctions or externally imposed symmetry breaking. Instead, it emerges intrinsically from the spatial asymmetry of the sample geometry and the vortex dynamics it induces. By analyzing the Cooper pair density $|\psi|^2$ across multiple layers of the meso-wedge, we show that vortex nucleation is highly sensitive to both geometry and the value of $\kappa$, resulting in nontrivial configurations that break inversion symmetry and differ between opposite current directions. The observed asymmetry in vortex configurations for $\mathbf{J} > 0$ and $\mathbf{J} < 0$ serves as a direct microscopic signature of the superconducting diode effect. In particular, vortex entry preferentially occurs near thinner regions of the sample, where reduced material thickness lowers the energy barrier, reinforcing the directional vortex motion under applied current.\\\\
Furthermore, we observe that increasing the magnetic field modifies the spatial distribution of vortices and suppresses $\gamma_{d}(\mathbf{H})$, indicating a competition between $\mathbf{H}$ strength and geometric confinement. Our results demonstrate that the superconducting diode effect can arise purely from vortex-mediated mechanisms and broken spatial symmetry, offering a novel and intrinsic route to nonreciprocal superconducting transport. These findings align closely with recent experimental observations based on SQUID-on-tip microscopy and critical $\mathbf{J}$ measurements, supporting the relevance and validity of the proposed theoretical framework. Overall, our work contributes to the fundamental understanding of superconducting rectification and may guide the design of future nonreciprocal superconducting devices without relying on complex hetero-structures or artificial junctions.
\section{ACKNOWLEDGMENTS}
C. Aguirre wants to thank S. Aguirre and M. Aguirre for useful discussions.  C. Aguirre thanks the CNPq grant number process: 174045/2023-9 for financial support.  J. Fáundez acknowledges the support from ANID Fondecyt grant number 3240320. Powered@NLHPC: This research was partially supported by the NLHPC supercomputing infrastructure (CCSS210001).
\section{DATA AVAILABILITY}\label{sec5}
The code used for the simulations was implemented in $\mathbf{Matlab}$\textsuperscript{\textregistered} and is available upon request from the corresponding author. The computational time required to obtain results depends on the sample size and simulation parameters. For the sample studied in this work, the average computation time was approximately 12 days, using a workstation equipped with an AMD Ryzen™ 9 7950X3D processor (32 threads), 64 GB of RAM, and running Ubuntu as the primary operating system.

\end{document}